\documentclass[12pt,preprint]{aastex}
\begin{document}
\title{Analytical Expressions for Spiral Arm Gravitational Potential and Density}
\author{ Donald P. Cox}
\affil{Department of Physics, University of Wisconsin-Madison, 1150 University Ave., Madison, WI 53706 USA}
\email{ cox@wisp.physics.wisc.edu}
\and
\author{ Gilberto C. G\'omez}
\affil{Department of Astronomy, University of Wisconsin-Madison, 475 N. Charter St., Madison, WI 53706 USA}
\email{ gomez@wisp.physics.wisc.edu}
\begin{abstract}
When modeling the three-dimensional hydrodynamics of interstellar material rotating in a galactic gravitational potential, it is useful to have an analytic expression for gravitational perturbations due to stellar spiral arms.  We present such an expression for which changes in the assumed characteristics of the arms can be made easily and the sensitivity of the hydrodynamics to those characteristics examined.  This analytic expression also makes it easy to rotate the force field at the pattern angular velocity with little overhead on the calculations.
\end{abstract}
\section{INTRODUCTION}
\label{section:Intro}
In this paper we present analytic expressions for the perturbation of the galactic axisymmetric gravitational potential due to redistribution of part of the stellar disk mass into spiral arms, and for the distribution of density responsible for that potential.  Adjustable parameters include the number of arms, $N$, the pitch angle, $\alpha$, the radial scale length of the dropoff in density amplitude of the arms, $R_{\rm s}$, the midplane arm density, $\rho_{\rm o}$ at fiducial radius $r_{\rm o}$, and the scale height of the stellar arm perturbation, $H$.

The amplitude of the spiral density distribution whose gravitational potential we set out to find is given by:
\begin{equation}
\rho_A(r, z) = \rho_{\rm o}  \exp\left(-\frac {r-r_{\rm o}}{R_{\rm s}}\right) {\rm sech}^2\left(\frac{z}{H}\right).
\label{equation:One} 
\end{equation}
Modulating this by a simple sinusoidal pattern in $\phi$, following a logarithmic spiral with a pitch angle $\alpha$, the overall density perturbation is
\begin{equation}
\rho(r, \phi, z) = \rho_A(r, z)\cos(\gamma)
\label{equation:Two} 
\end{equation}
where
\begin{equation}
\gamma = N \left[\phi - \phi_p(r_{\rm o}) - \frac {\ln(r/r_{\rm o})}{\tan(\alpha)}\right].
\label{equation:Three} 
\end{equation}

More complicated azimuthal arm structures can be constructed with linear combinations of these solutions of the form
\begin{equation}
\rho(r, \phi, z) = \rho_A(r, z) \sum_n {\rm C_n}\cos( n \gamma).	
\label{equation:Four} 
\end{equation}
In a particularly interesting example, the density behaves approximately as a cosine squared in the arms but is separated by a flat interarm region occupying half the volume.  It has three terms in its sum, with $C_1 = 8/(3 \pi) $, $C_2 = 1/2$, and $C_3 = 8/(15 \pi)$.  The resulting phase pattern is compared with that of a simple sinusoid in Figure~\ref{fig1}.
%
%
%\begin{figure}
%\caption{ Comparison of Modulation Functions for Sinusoidal and Concentrated Arms.}
%\label{fig1}
%\end{figure}

An important feature of such a perturbation is that its average density is zero.  Its potential can thus be added to observationally constrained models for the azimuthally averaged potential without altering the latter.  In addition, because the assumed arm perturbation extends over all radii, it is important that the average density be zero at both large and small radii where the arms do not actually exist.  The radial exponential damping introduced in Equation~\ref{equation:One} is also useful in this regard.  Thus the gradient of the perturbation potential in the calculation region is provided predominantly by the local distribution of material.

\section{THE POTENTIAL}
\label{section:Potent}
In developing an expression for the potential, we first considered an infinite sinusoidally oscillating density pattern in a rectilinear coordinate system.  Mathematically it was equivalent to Equations~\ref{equation:One} and ~\ref{equation:Two} with $R_{\rm s}$\ infinite, $\gamma$\ replaced with $kx$, and $y$ extending from + to - $\infty$, normal to the wave vector.  For this density distribution, we evaluated the potential numerically and fitted the results versus phase ($kx$) and $z$\ with a simple functional form.  We then remapped the solution's phase into the desired spiral pattern and reintroduced the radial dropoff factor, $\exp(-\frac {r-r_{\rm o}}{R_{\rm s}}$), obtaining a trial potential function.

The resulting potential has three functional parameters dependent on radius:
\begin{eqnarray}
	K_{\rm n} &  = & \frac {n N}{r \sin(\alpha)} \\
	\beta_{\rm n} &  = & K_{\rm n} H (1+ 0.4 K_{\rm n} H) \\
	D_{\rm n} &  = & \frac {1 + K_{\rm n} H + 0.3 (K_{\rm n} H)^2}{1+0.3 K_{\rm n} H}. 
\label{equation:Params} 
\end{eqnarray}
With these parameters, the trial potential corresponding approximately to the density distribution of Equation~\ref{equation:Four} is
\begin{equation}
\Phi(r, \phi, z) = - 4 \pi G H \rho_{\rm o}  \exp\left(-\frac {r-r_{\rm o}}{R_{\rm s}}\right) 
\sum_n \left(\frac {{\rm C_n}}{K_{\rm n} D_{\rm n}}\right)\cos( n \gamma) \left[{\rm sech}\left(\frac {K_{\rm n} z}{\beta_{\rm n}}\right) \right] ^{\beta_{\rm n}}
\label{equation:Eight} 
\end{equation}

\section{THE DENSITY}
\label{section:Dense}
Our next step was just an inversion of attitude, from considering the density function of equations~\ref{equation:One} through ~\ref{equation:Four} as primary and the potential function of equation~\ref{equation:Eight} as an approximate solution, to the inverse perspective.  Equation~\ref{equation:Eight} is our final equation for the potential perturbation, and by evaluating
\begin{equation}
\rho = \frac {\nabla^2 \Phi}{4 \pi G},
\label{equation:Nine} 
\end{equation}
we find the exact corresponding density function.
Because of the radial dependence of the parameters $K_{\rm n}$, $\beta_{\rm n}$, and $D_{\rm n}$, the radial derivatives of the Laplacian are particularly messy.  The full solution is given in the Appendix; for each $n$\ there is one dominant term for small $H/r$\ and not too large $z/H$\ such that:
\begin{equation}
\rho (r, \phi, z) \approx \rho_{\rm o}  \exp\left(-\frac {r-r_{\rm o}}{R_{\rm s}}\right)
\sum_n {\rm C_n}
\left[\frac {K_{\rm n} H }{D_{\rm n}}\frac {\beta_{\rm n} +1}{\beta_{\rm n}}\right]
\cos(n \gamma) 
\left[{\rm sech}\left(\frac {K_{\rm n} z}{\beta_{\rm n}}\right) \right] ^{(2+\beta_{\rm n})}.	
\label{equation:Ten} 
\end{equation}
Of course, for particular choices of parameters, the Laplacian can be evaluated numerically to determine the corresponding density function.  We did so to check the exact density solution of the Appendix, for example. 

\section{EXAMPLES}
\label{section:Exam}
\begin{figure}
%\epsscale{0.75}
\plotone{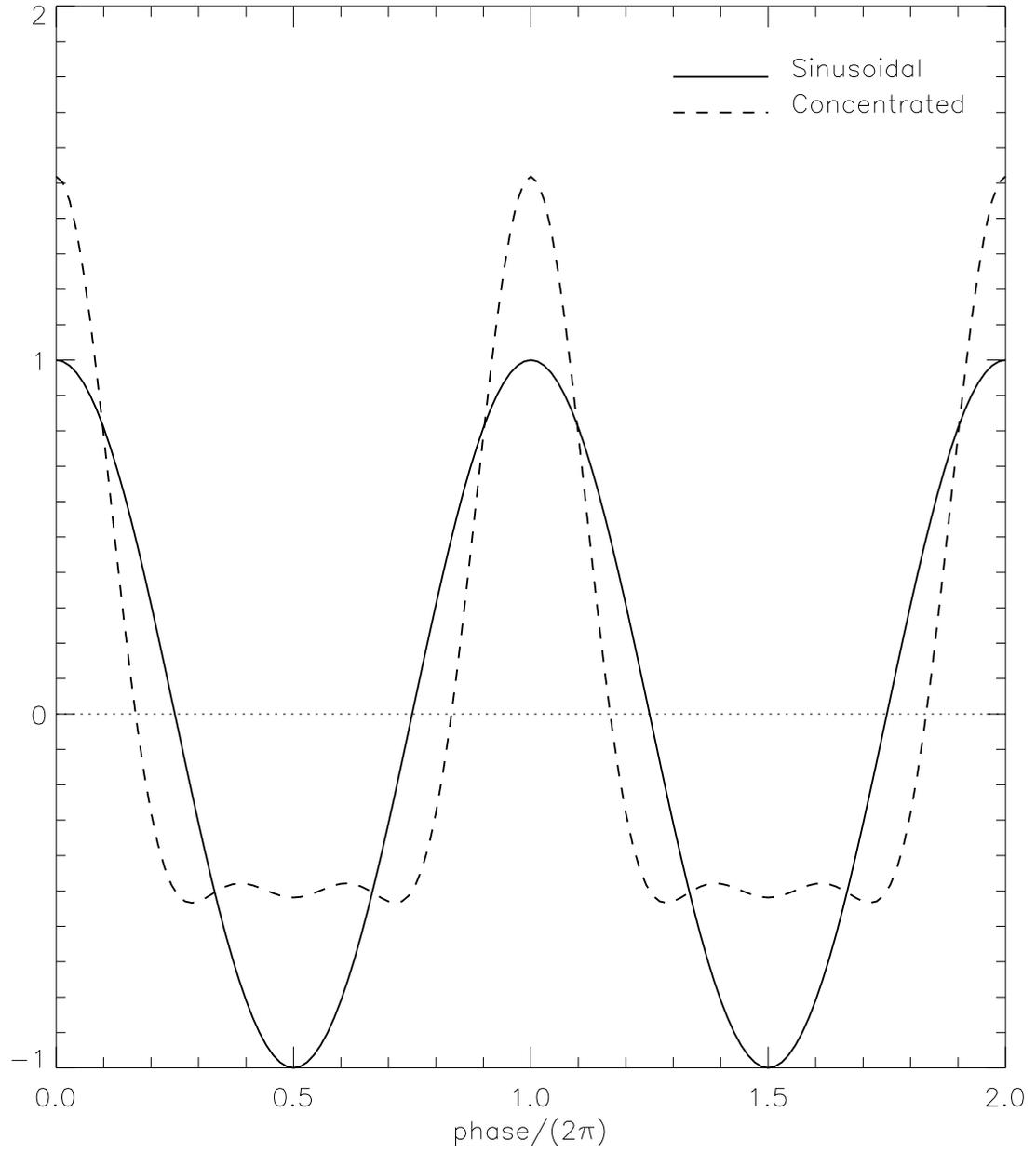}
\epsscale{1.00}
\caption{ Comparison of Modulation Functions for Sinusoidal and Concentrated Arms.}
\label{fig1}
\end{figure}

We next present the results for two cases, one with simple sinusoidal arms and one with the more concentrated arms, as per Figure~\ref{fig1}.  The parameters in both cases are $N$\ = 2, $\alpha = 15^{\circ}$, $R_{\rm s}$ = 7 kpc, $\rho_{\rm o} = m n_{\rm o}$, $n_{\rm o}$ = 1 atom $cm^{-3}$\ at $r_{\rm o}$ = 8 kpc, and $H$ = 0.18 kpc (chosen to match the scaleheight of the thin stellar disk of Dehnen and Binney, 1998, Model 2).  The average mass per atom is $m = (14/11) m_H$.  Unless otherwise noted, results are shown versus $r$\ and $z$\ along a radial cut passing through maximum arm density at $r$\ = 8 kpc.

\begin{figure}
\epsscale{0.75}
\plotone{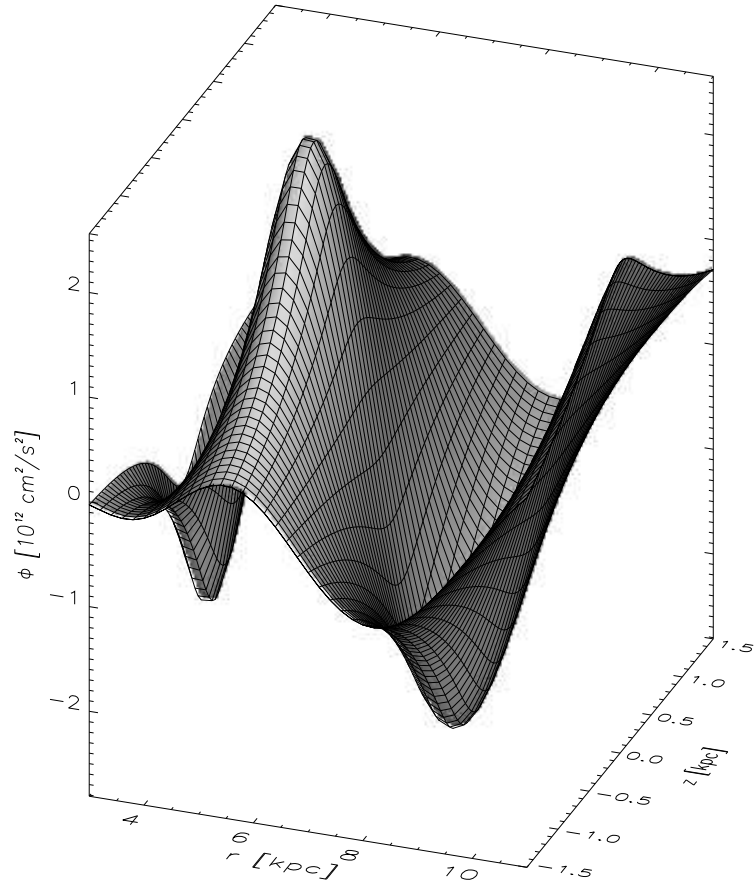}
\epsscale{1.00}
\caption{Gravitational Potential for Sinusoidal Arms. This plot and all subsequent ones versus $r$\ and $z$\ are for a radial cut at constant $\phi$\ which intersects an arm at 8 kpc. Parameters are given in Section \ref{section:Exam}.}
\label{fig2}
\end{figure}

Figures~\ref{fig2} and ~\ref{fig3} show the perturbing gravitational potential and corresponding density found from the Laplacian of the potential, for the sinusoidal arm case.  Figures~\ref{fig4} and \ref{fig5} are similar but for the concentrated arms case.  In both cases, the density functions are almost indistinguishable from the assumed function of Equation~\ref{equation:Four} or the dominant term form of Equation~\ref{equation:Ten}; the differences are much smaller than the uncertainty in how the true arms should be represented.  As a result, Equation~\ref{equation:Four} (or \ref{equation:Ten}) can be used with considerable confidence as an approximation to the density distribution responsible for the potential.
%
%\begin{figure}
%\caption{Gravitational Potential for Sinusoidal Arms.}
%\label{fig2}
%\end{figure}
%
%\begin{figure}
%\caption{Corresponding Density Distribution for Sinusoidal Arms.}
%\label{fig3}
%\end{figure}
%
%\begin{figure}
%\caption{ Gravitational Potential for Concentrated Arms.}
%\label{fig4}
%\end{figure}
%
%\begin{figure}
%\caption{Corresponding Density Distribution for Concentrated Arms.}
%\label{fig5}
%\end{figure}
%
  
It is somewhat easier to compare these two cases by examining separately their radial and vertical behaviors.  Figures~\ref{fig6} and \ref{fig7} compare the midplane densities and potentials versus radius.  The maxima and minima of the density functions are bounded by the decaying exponential envelope with $R_{\rm s}$\ = 7 kpc.  This value for the radial dropoff scale was chosen so that the amplitude of the potential variation would not depend strongly on radius.  Our ``sinusoidal" case is sinusoidal with phase (and in $\phi$), but in the logarithmic spiral, the effective wavelength is proportional to radius.  The ``concentrated" case has the desired flatter interarm density and sharper peaks at the arms.  The difference is moderated somewhat in the potential function, but the flatter peaks (interarm) and sharper valleys of the arms are apparent in the concentrated arm case.  Figures~\ref{fig8} and \ref{fig9} compare the vertical structures at the location of an arm.  The greater density and deeper and sharper potential well of the concentrated case are evident.  

Comparison of Figures~\ref{fig8} and \ref{fig9} emphasizes the much greater scale height of the potentials, compared to that of the responsible densities, evident also in Figures~\ref{fig2} through \ref{fig5}.  The potential scale height depends on both the density scale height and on the radial wavelength.  For an infinite disk (infinite radial wavelength), the vertical scale height is also infinite, having a constant gradient outside the mass distribution.  Our results successfully model the useful cases with $K_{\rm n} H$\ small to moderate. 
%
%\begin{figure}
%\caption{Midplane Densities of the Two Cases, Versus Radius.}
%\label{fig6}
%\end{figure}
%
%\begin{figure}
%\caption{Midplane Potentials of the Two Cases, Versus Radius.}
%\label{fig7}
%\end{figure}
%
%\begin{figure}
%\caption{Densities of the Two Cases at r = 8 kpc, Versus Height. }
%\label{fig8}
%\end{figure}
%
%\begin{figure}
%\caption{Potentials of the Two Cases at r = 8 kpc, Versus Height.}
%\label{fig9}
%\end{figure}

%A more intuitive view of the potential distribution is offered by Figure~\ref{fig10}, which shows the 
%spiral pattern of the midplane perturbation potential of the concentrated arm case.  Combined with 
%Figure~\ref{fig4} which shows the radial and vertical distributions on one radial cut, it provides a 
%full sense of the behavior.  The pattern for the sinusoidal case is quite similar; the present case 
%differentiated only by the somewhat flatter peaks and sharper valleys.
%
%\begin{figure}
%\caption{Midplane Perturbation Potential for the Concentrated Arm Case.  The domain is 22 kpc on %a side.  The rapidly oscillating potential in the inner 3 kpc is not shown.}
%\label{fig10}
%\end{figure}

The impressions invoked by the density distributions of Figures~\ref{fig3} and \ref{fig5} can be somewhat misleading.  These densities must be regarded as perturbations to an azimuthally uniform stellar disk with the same vertical scale height.  In Figure~\ref{fig11}, a disk component with the same radial dropoff and scale height as the perturbation, and just sufficient amplitude to make the net density  everywhere positive, has been added to the perturbation density.  In the figure captions, this disk is referred to as the Net Mass Disk, as it provides the material for the arms.  This picture provides a better idea of the assumed density structure of the arms.  

Those tempted to find the potential of the above combination of disk plus perturbation may wish to consider an alternative approach to evaluation of the spiral arm potential that has been pursued by \citet{Bbara}; they specify a positive density pattern and evaluate the potential with mixed analytic and numerical techniques.
%
%\begin{figure}
%\caption{Density Distribution of Concentrated Arm Perturbation Plus Net Mass Disk.}
%\label{fig11}
%\end{figure}

Having created this intuitive view of the density distribution, it is used in Figure~\ref{fig12} to illustrate the qualitative distribution of arm material in the galactic midplane.
%, responsible for the potential distribution in Figure~\ref{fig10}.
%
%\begin{figure}
%\caption{Midplane Arm Density for the Concentrated Arm Case.  The domain is 22 kpc on a side.  %The rapidly oscillating density in the inner 3 kpc is not shown.  Density includes net mass disk with %the same scale height, radial dropoff rate, and total mass as the arms.}
%\label{fig12}
%\end{figure}

In Figures~\ref{fig13} and \ref{fig14}, various amounts of perturbation density are shown added to a representative full stellar disk.  The disk parameters were chosen to approximate the thin stellar disk in Model 2 of \citet{D&B}.  It has a ${\rm sech}^2(z/H) $\ form with a scale height of 0.18 kpc, a radial dropoff scale of 2.4 kpc, and an atomic number density of 3.21 cm$^{-3}$\ at $r$\ = 8 kpc.  In the sinusoidal arm case, the added perturbation has, at 8 kpc, an amplitude equal to the fractions of the total 3.21 cm$^{-3}$\ disk density that are shown in the legend.  A fraction of 0.4, for example implies total interarm and arm densities of 0.6 and 1.4 times the unperturbed density.  At smaller radii, the perturbation is a smaller fraction of the full disk because of their different radial dropoff scales.\footnote{The extreme difference between the 2.4 kpc radial dropoff scale for the average thin disk density and our choice of 7 kpc for the radial dropoff scale of the perturbation density was not guided by observations.  We are using the perturbation in 3D MHD modeling of the response of the interstellar medium to the stellar arm potential, and did not want to introduce, at least initially, a strong radial gradient in the depth of the arm potential well.}  In the concentrated arm case, the normalization is slightly different to provide similar arm to interarm density difference at fixed radius, as per Figure~\ref{fig1}.  In this case, a 0.4 perturbation fraction implies total interarm and arm densities of 0.8 and 1.6 times the unperturbed density at $r$\ = 8 kpc.  (The average density is the same in the two cases because the concentrated arms are narrower.)  These figures provide a sense of the degree of modulation required for an observable density contrast in the presence of the steep gradient of the unperturbed disk. Rather large perturbations are required just to flatten the density profile.  When the perturbation is large enough to give the total density a local maximum, that maximum is shifted radially inward from the perturbation peak.  

%
%%\begin{figure}
%\caption{Midplane Density of Disk+Arms for Various Arm Amplitudes, Sinusoidal Arm Case.}
%\label{fig13}
%\end{figure}
%
%\begin{figure}
%\caption{Midplane Density Versus Radius of Disk+Arms for Various Arm Amplitudes, %Concentrated Arm Case.}
%\label{fig14}
%\end{figure}

Figures~\ref{fig15} and \ref{fig16} further examine the stellar density distribution.  Figure~\ref{fig15} shows the vertical distribution of the density versus radius for the stellar disk described above plus ``0.57" fraction sinusoidal arms. Figure~\ref{fig16} is similar, but the perturbation is shown added to the entire axisymmetric galactic density distribution of Model 2 of \citet{D&B}.  (Actually, the potentials were added and then the density evaluated by taking the Laplacian numerically.)  In the spiral arm modeling of \citet{G&C}, these are the parameters used for the two-armed case.  It had been our prejudice that a 57\% modulation of the disk density at $r$\ = 8 kpc represented a rather extreme situation, and we were therefore surprised to learn its rather modest effect on the density structure in comparison to the radial gradient.

%This last point is emphasized further by Figure~\ref{fig17}, which shows the midplane density 
%distribution of the stellar disk plus ``0.2" fraction concentrated arms.  Because of the constant 
%vertical scale height, the projected stellar surface density is proportional to this midplane density.  
%Thus, Figure~\ref{fig17} also represents the anticipated red light distribution of main sequence 
%stars for a galaxy viewed face-on.  The distribution is plotted logarithmically, with 5 contours per 
%decade, and therefore 2 contours per magnitude of surface brightness.  In this example, the arms 
%become noticeable as local maxima only very far out in the galaxy where the more gradual dropoff 
%of the arm amplitude enhances the contrast.  In the neighborhood of 8 kpc radius, the arms at this 
%amplitude provide only a local flattening of the surface brightness profile.

These results can be compared with measurements of the actual red light profiles of spiral galaxies, if the contamination due to red supergiants can be neglected, to provide a reasonable estimate for the arm perturbation amplitude.  This was done, for example, for M51 by \citet{R&R} and for several other galaxies by \citet{R&Z}.  They concluded that in galaxies with strong arms, the arm to interarm stellar density contrast was roughly a factor of two.  They further found that in some two-armed spirals, the non-axisymmetric structure was dominated by the fundamental sinusoidal component, implying very broad arms (as in our ``sinusoidal" case); but in cases with very strong arms there was a significant higher harmonic, implying narrower arms (as in our ``concentrated" case).  Thus, density contrasts at the upper end of those we have explored seem to be relevant to at least strong armed galaxies, as do both varieties of the arm modulation functions of Figure 1. 
%
%\begin{figure}
%\caption{Density Distribution of Disk Plus Arms for 0.4  Amplitude, Sinusoidal Arm Case.}
%\label{fig15}
%\end{figure}
%
%\begin{figure}
%\caption{The Whole Nine Yards.  Figure shows the full density represented by Model 2 of %\citet{D&B} plus a sinusoidal arm perturbation with our standard parameters (see text, Examples) %with an amplitude at 8 kpc of 40\% of the thin disk density.  This plot versus $r$\ and $z$\ is for a %radial cut at constant $\phi$\ for which the perturbation has a maximum density at 8 kpc. This is the %two-armed spiral model used by \citet{G&C} to explore the gaseous response in MHD.  Note that %the perceptible density contrast is small even at this amplitude.}
%\label{fig16}
%\end{figure}
%
%
%\begin{figure}
%\caption{Logarithm of Midplane Density of Disk Plus Arms for 0.2 Amplitude Concentrated Arm %Case.  The projected surface density of disk stars is proportional to this midplane density.  There %are five contours per decade.}
%\label{fig17}
%\end{figure}
%

\section{ROTATING THE PATTERN}
\label{section:Rotor}
If one is calculating the hydrodynamics in the inertial frame, it is necessary to rotate the perturbing potential at the pattern angular velocity $\Omega$.  For $\phi$\ increasing counterclockwise and a clockwise rotation, examination of Equations~\ref{equation:Three} and \ref{equation:Eight} reveals that each term in the sum consists of an amplitude function
\begin{equation}
A_n(r, z) =  - 4 \pi G H \rho_{\rm o}  \exp\left(-\frac {r-r_{\rm o}}{R_{\rm s}}\right) \left(\frac {{\rm C_n}}{K_{\rm n} D_{\rm n}}\right) \left[{\rm sech} \left( \frac {K_{\rm n} z}{\beta_{\rm n}} \right) \right] ^{\beta_{\rm n}},.
\label{equation:Eleven} 
\end{equation}
multiplied by a phase function $\cos(n\Gamma)$, where
\begin{equation}
\Gamma(r, \phi, t) = \gamma + N \Omega t = N \left[\phi +\Omega t - \phi_p(r_{\rm o}) - \frac {\ln(r/r_{\rm o})}{\tan(\alpha)}\right].
\label{equation:Twelve} 
\end{equation}
Thus, for each term in the sum of Equation~\ref{equation:Eight}, only the function $A_n(r, z)$\ needs to be evaluated and stored. 
At any time, they can be combined according to
\begin{equation}
 \Phi(r, \phi, z, t) =  \sum_n A_n(r, z) \cos\left(n \Gamma(r, \phi, t)\right) 
\label{equation:Thirteen} 
\end{equation}
%		
%
%\section{Discussion/Conclusions}
%\label{section:Concl}
%
%Words of wisdom?  

\acknowledgements

We are grateful for useful conversations with Bob Benjamin and Matt Bershady, 
to the NASA Astrophysics Theory Program for financial support under NASA Grant NAG 5-8417 to the University of Wisconsin-Madison, and to M\'exico's Consejo Nacional de Ciencia y Tecnolog\'{\i}a for support to G. C. G.

\appendix
\section{THE EXACT DENSITY DISTRIBUTION}

\begin{figure}
\epsscale{0.75}
\plotone{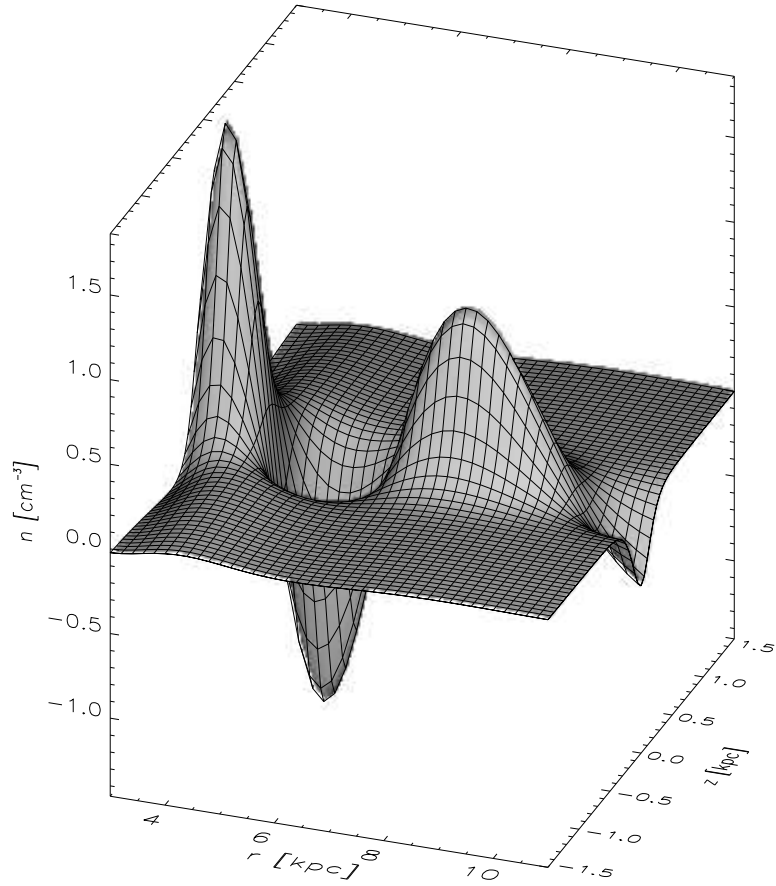}
\epsscale{1.00}
\caption{Distribution of Density Perturbation for Sinusoidal Arms.  Corresponding potential is shown in Figure \ref{fig2}.}
\label{fig3}
\end{figure}

\begin{figure}
\epsscale{0.75}
\plotone{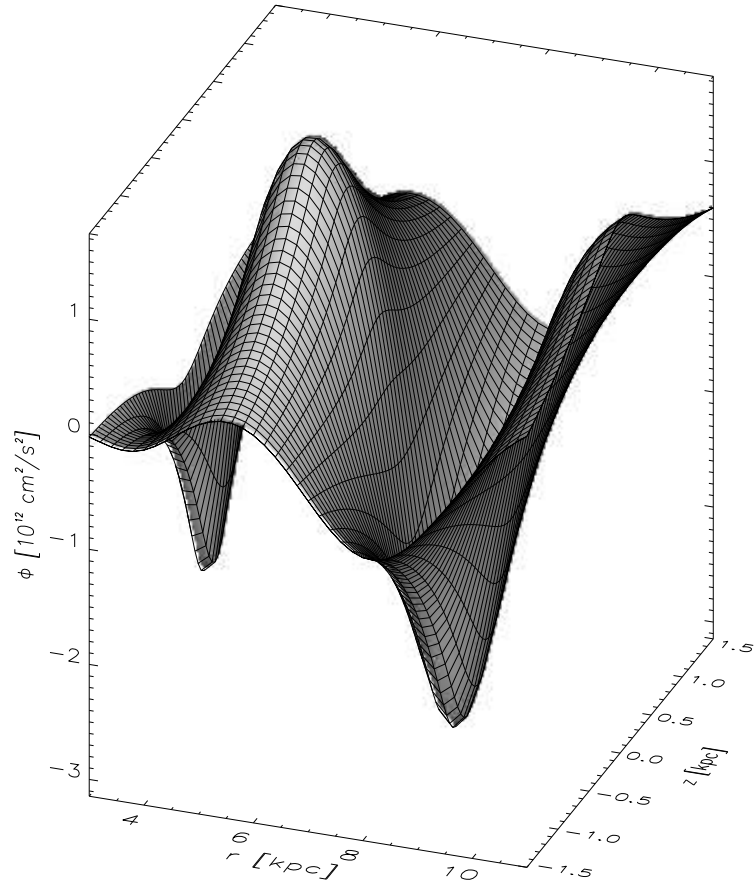}
\epsscale{1.00}
\caption{ Gravitational Potential for Concentrated Arms.}
\label{fig4}
\end{figure}

\begin{figure}
\epsscale{0.75}
\plotone{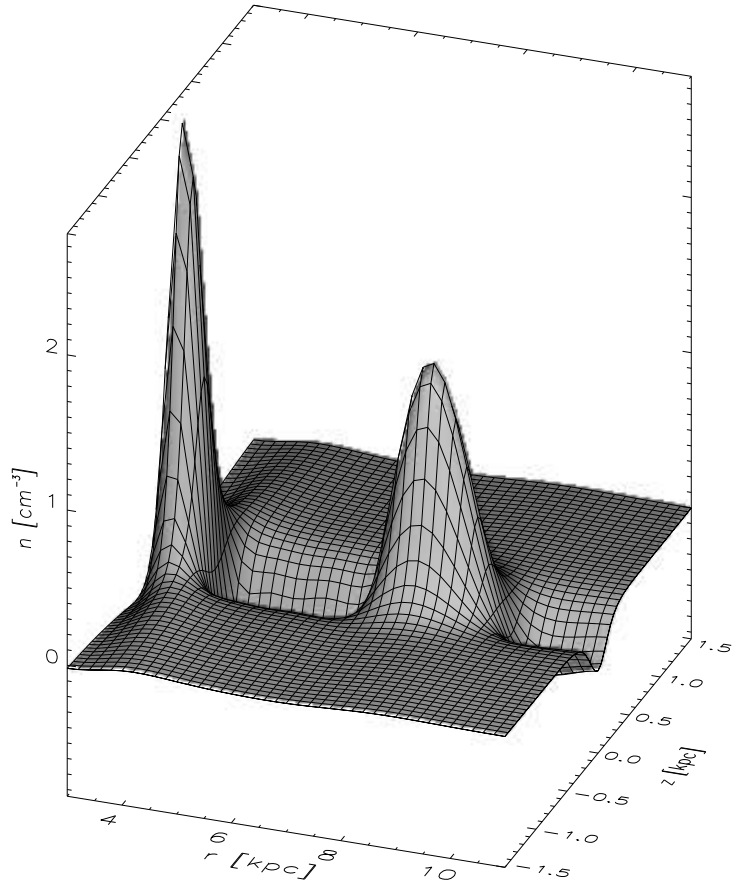}
\epsscale{1.00}
\caption{ Distribution of Density Perturbation for Concentrated Arms.  Corresponding potential is shown in Figure \ref{fig4}.}
\label{fig5}
\end{figure}

\begin{figure}
%\epsscale{0.75}
\plotone{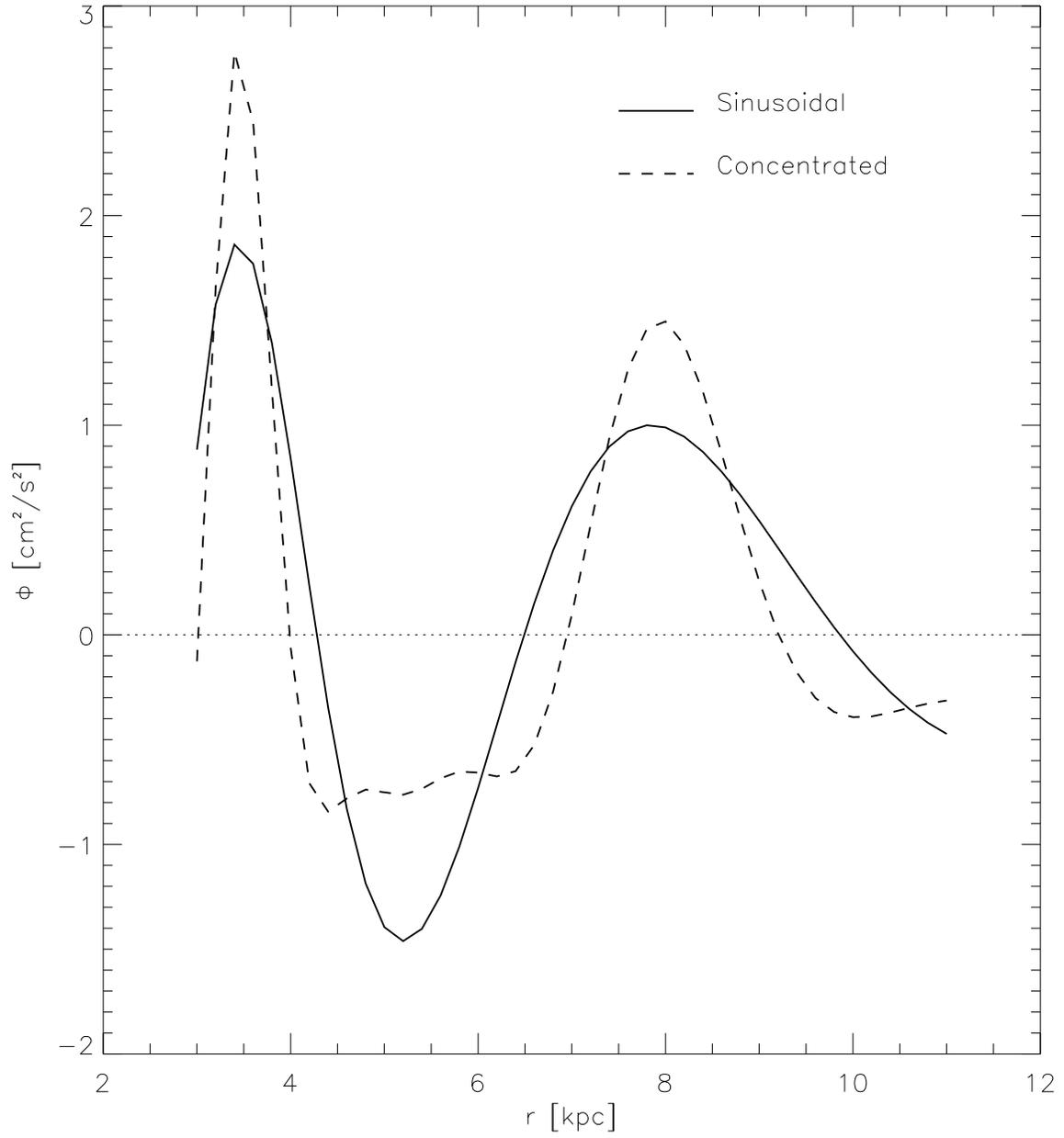}
\epsscale{1.00}
\caption{Midplane Densities of the Two Cases, Versus Radius.}
\label{fig6}
\end{figure}

\begin{figure}
%\epsscale{0.80}
\plotone{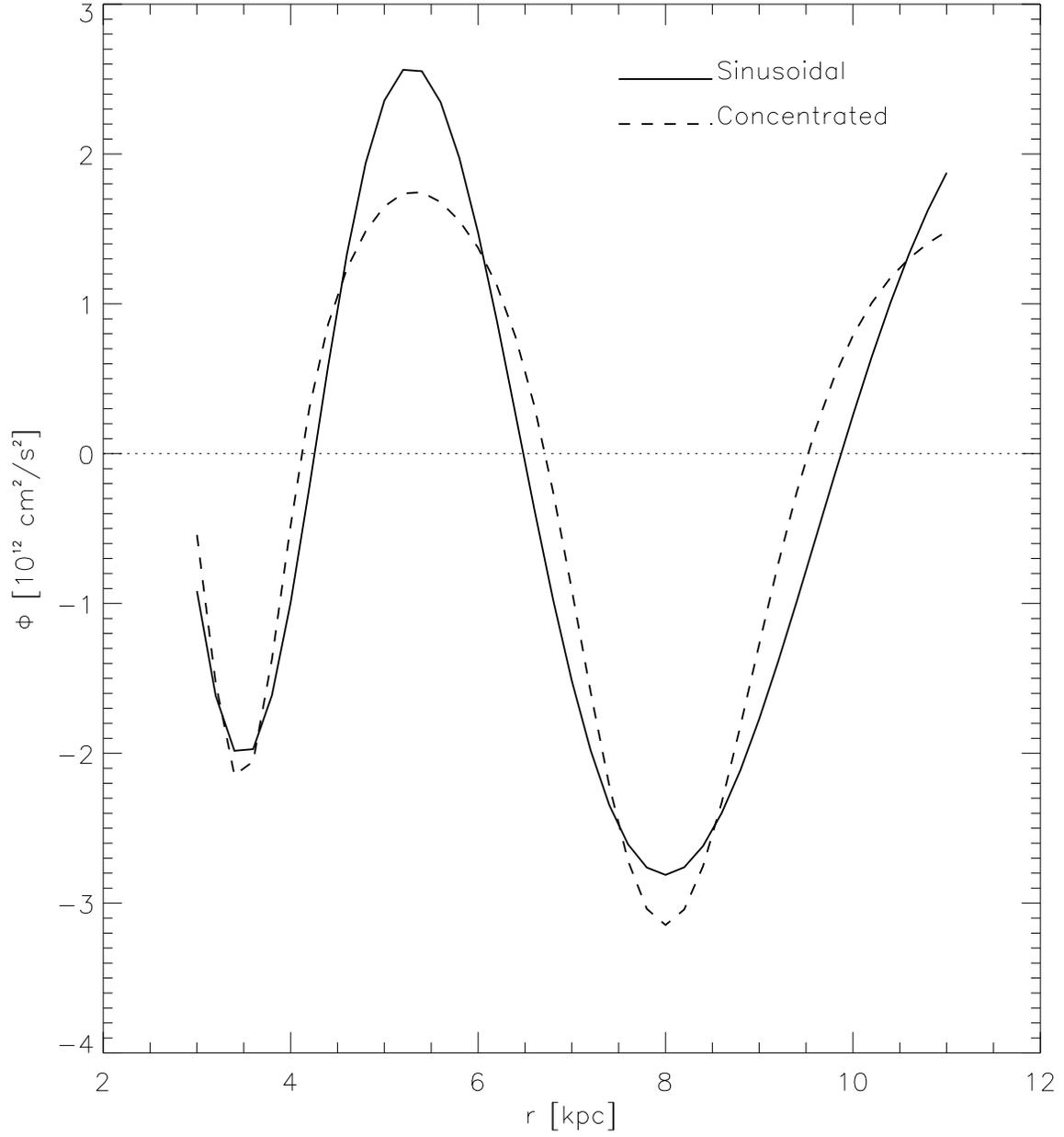}
\epsscale{1.00}
\caption{Midplane Potentials of the Two Cases, Versus Radius.}
\label{fig7}
\end{figure}

\begin{figure}
%\epsscale{0.85}
\plotone{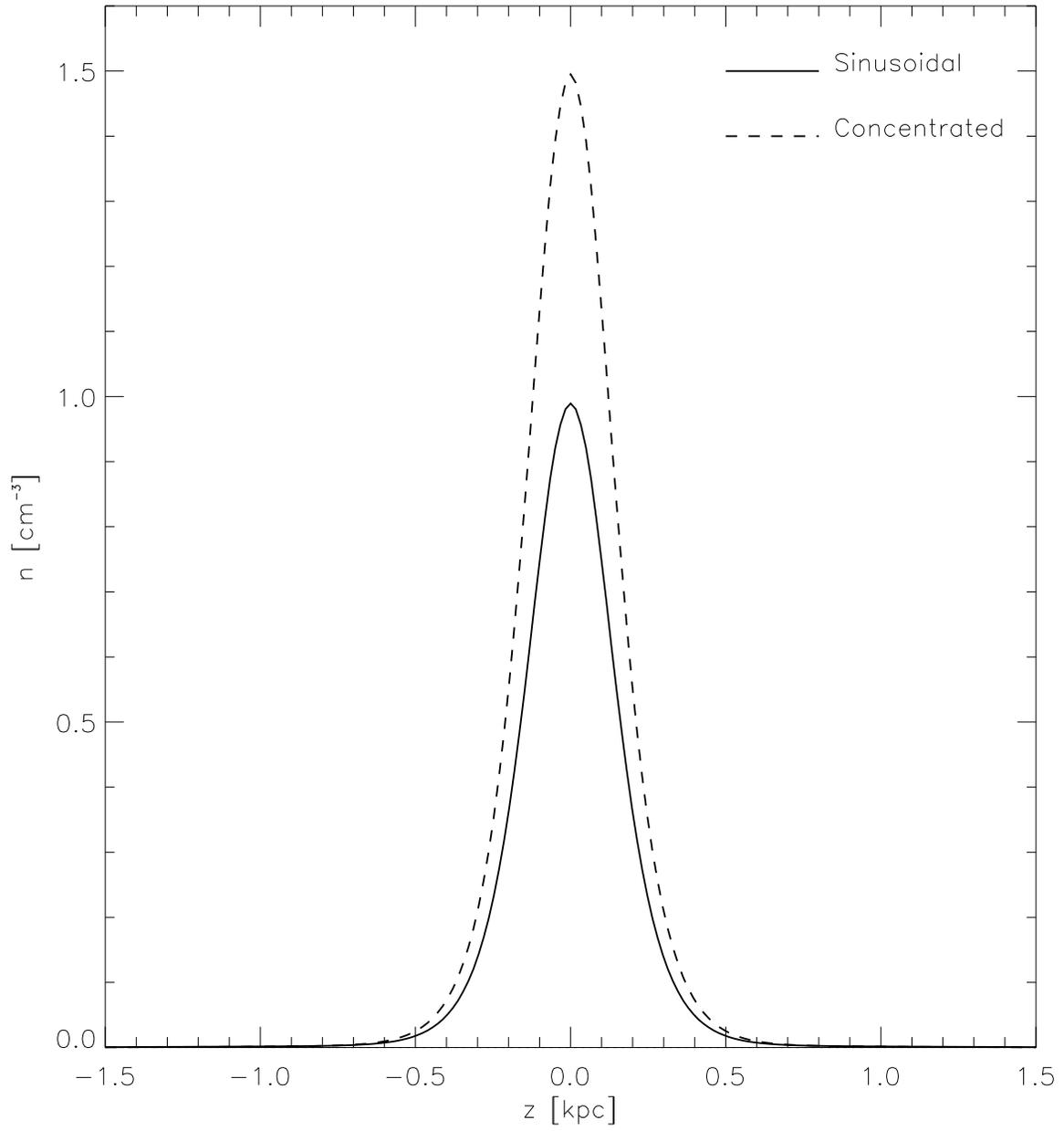}
\epsscale{1.00}
\caption{Densities of the Two Cases at r = 8 kpc, Versus Height. }
\label{fig8}
\end{figure}

\begin{figure}
%\epsscale{0.85}
\plotone{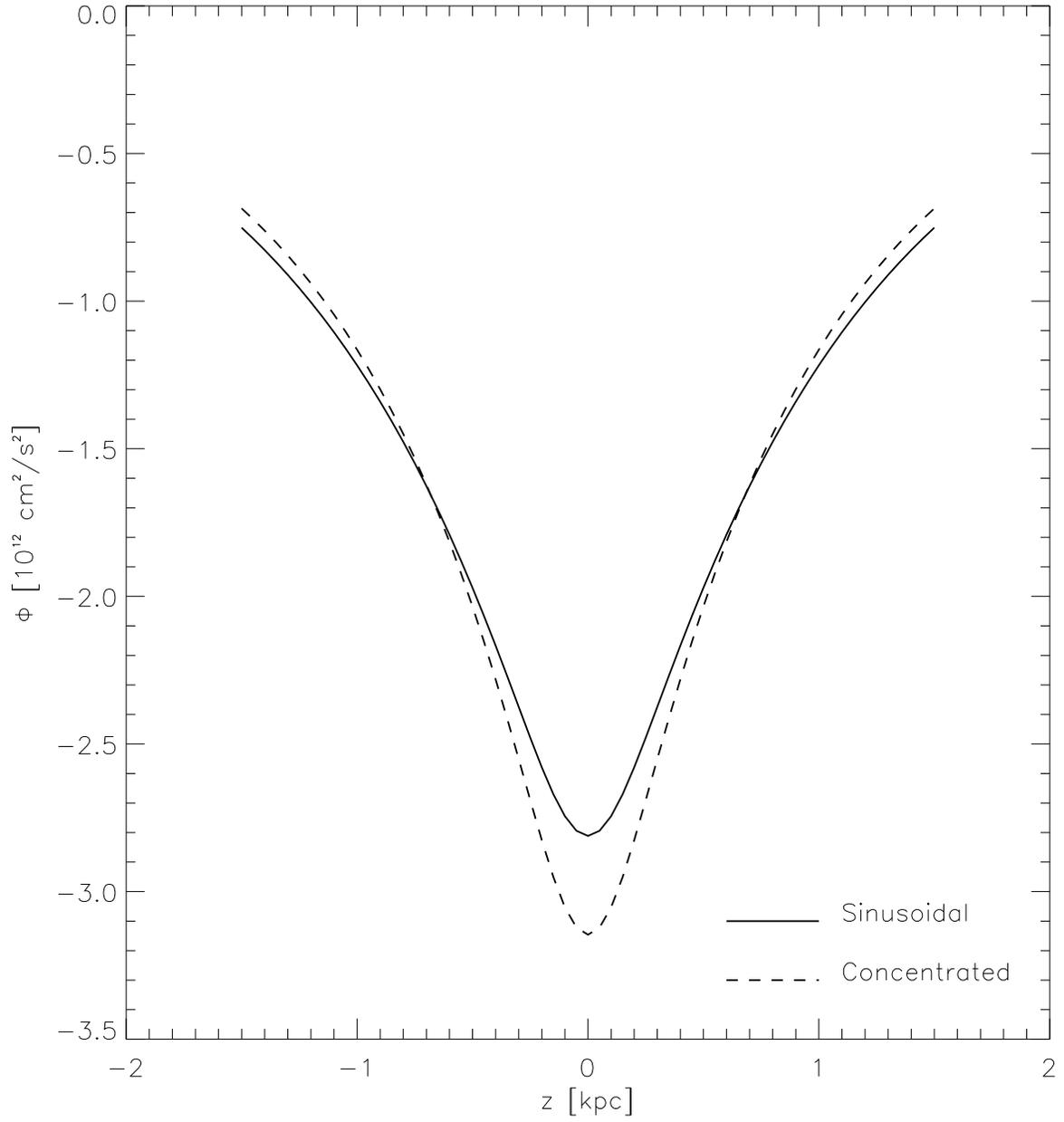}
\epsscale{1.00}
\caption{Potentials of the Two Cases at r = 8 kpc, Versus Height.}
\label{fig9}
\end{figure}

\begin{figure}
\epsscale{0.75}
\plotone{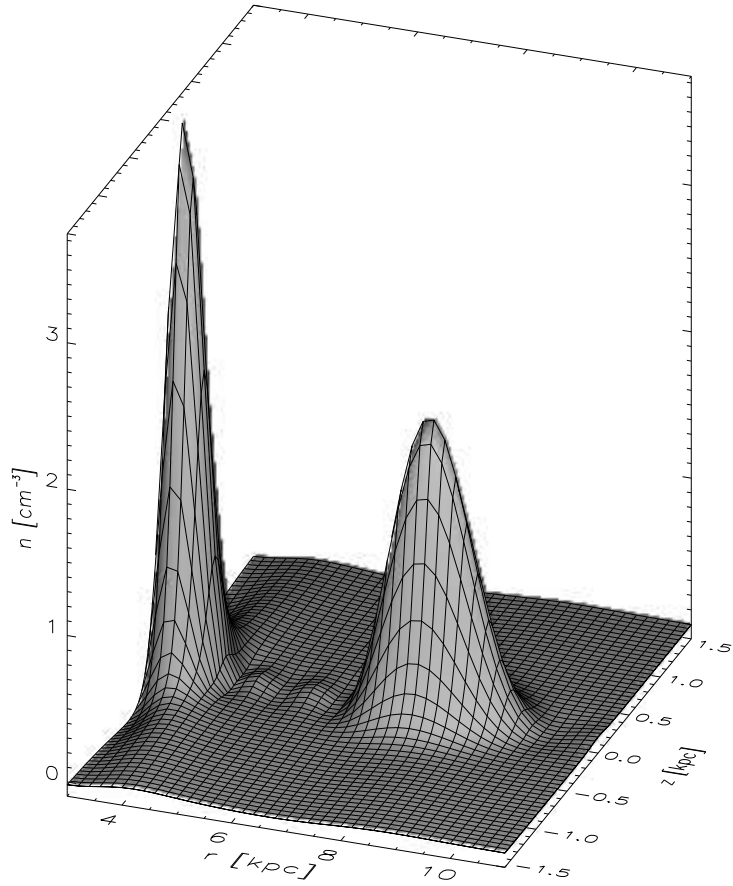}
\epsscale{1.00}
\caption{Density Distribution of Concentrated Arm Perturbation Plus Net Mass Disk.  See text for description of this intuitive distribution of the arm density, rather than the arm density perturbation, which has average density zero.}
\label{fig11}
\end{figure}

\begin{figure}
\epsscale{0.60}
\plotone{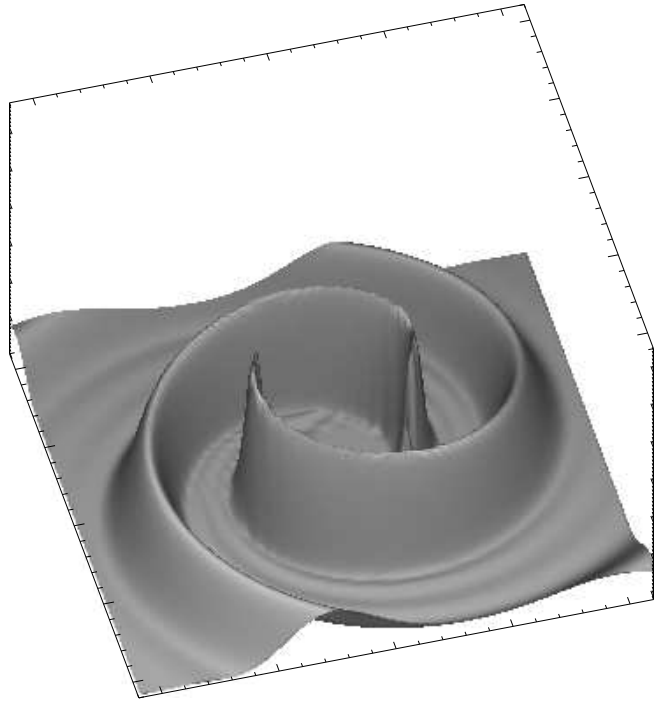}
\epsscale{1.00}
\caption{Midplane Arm Density for the Concentrated Arm Case.  The domain is 22 kpc on a side.  The rapidly oscillating density in the inner 3 kpc is not shown.  Density includes net mass disk with the same scale height, radial dropoff rate, and total mass as the arms.}
\label{fig12}
\end{figure}

\begin{figure}
\epsscale{0.75}
\plotone{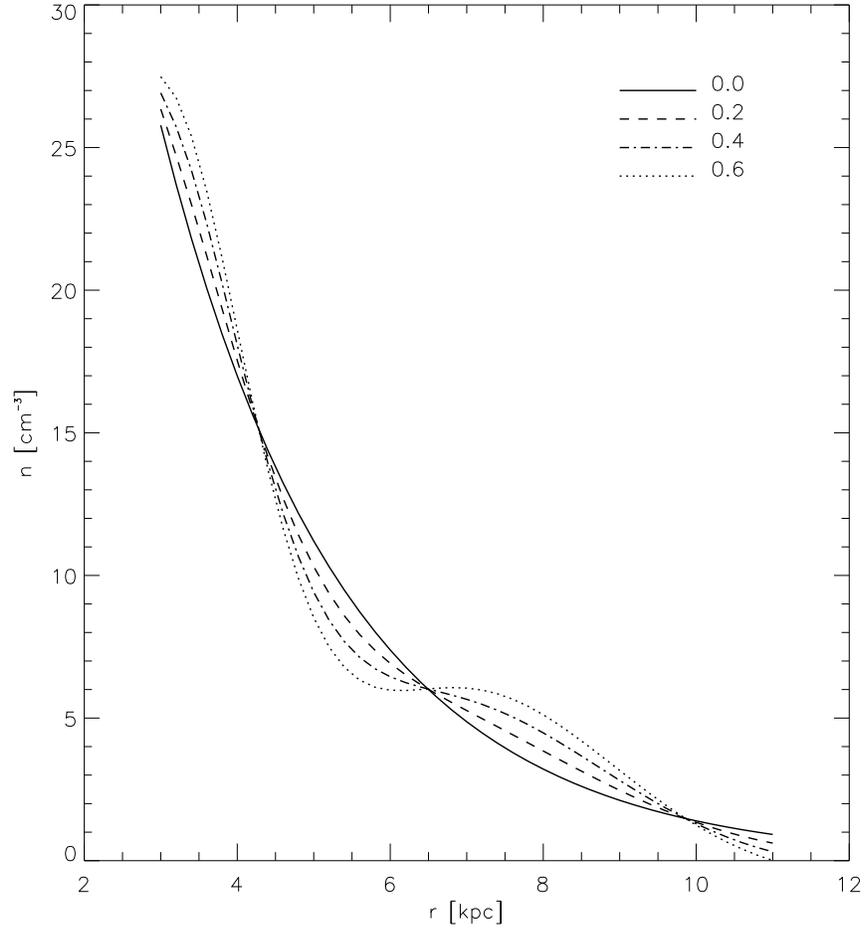}
\epsscale{1.00}
\caption{Midplane Density of Disk+Arms for Various Arm Amplitudes, Sinusoidal Arm Case.  The legend shows the fractional amplitude relative to the unperturbed disk at a radius of 8 kpc.}
\label{fig13}
\end{figure}

\begin{figure}
\epsscale{0.75}
\plotone{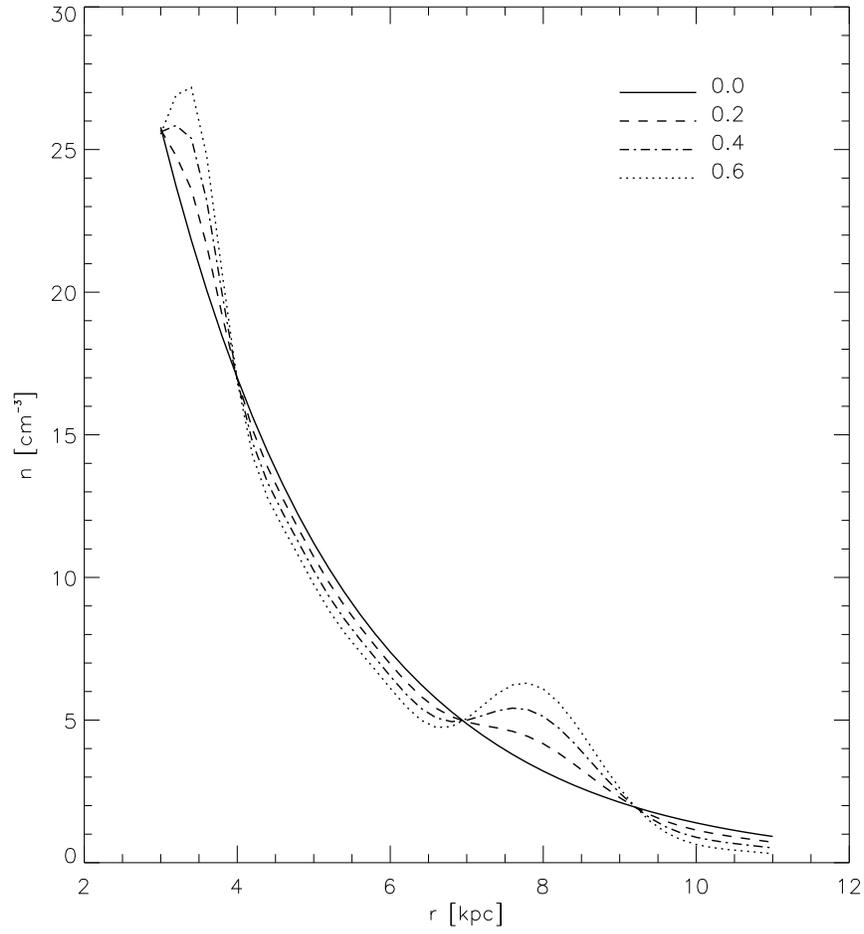}
\epsscale{1.00}
\caption{Midplane Density Versus Radius of Disk+Arms for Various Arm Amplitudes, Concentrated Arm Case.  See text for amplitude description.}
\label{fig14}
\end{figure}

\begin{figure}
\epsscale{0.75}
\plotone{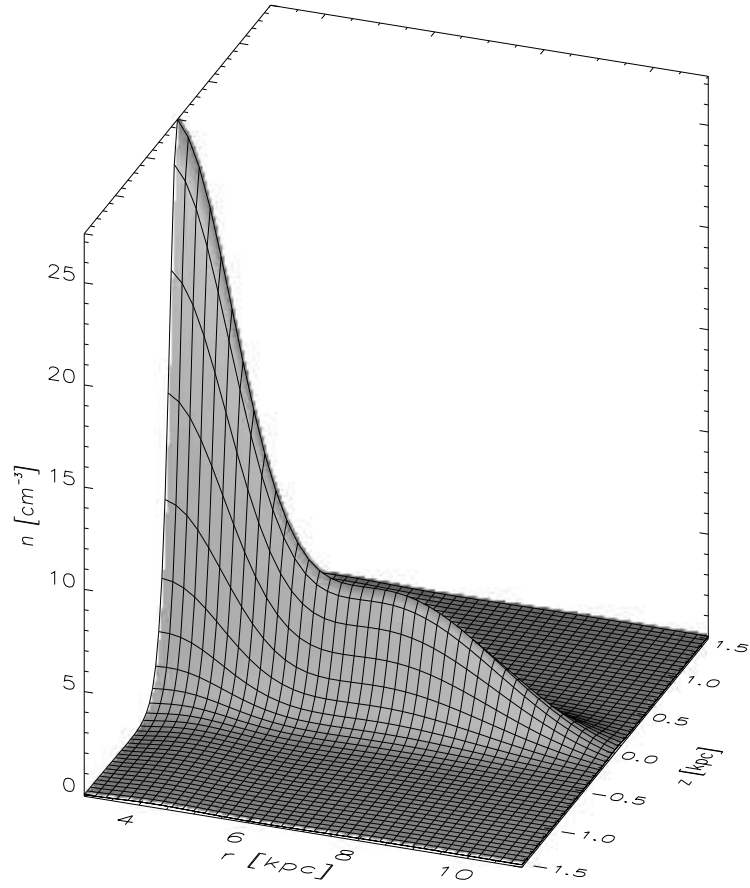}
\epsscale{1.00}
\caption{Density Distribution of Disk Plus Arms for 0.57  Amplitude, Sinusoidal Arm Case.}
\label{fig15}
\end{figure}

\begin{figure}
\epsscale{0.75}
\plotone{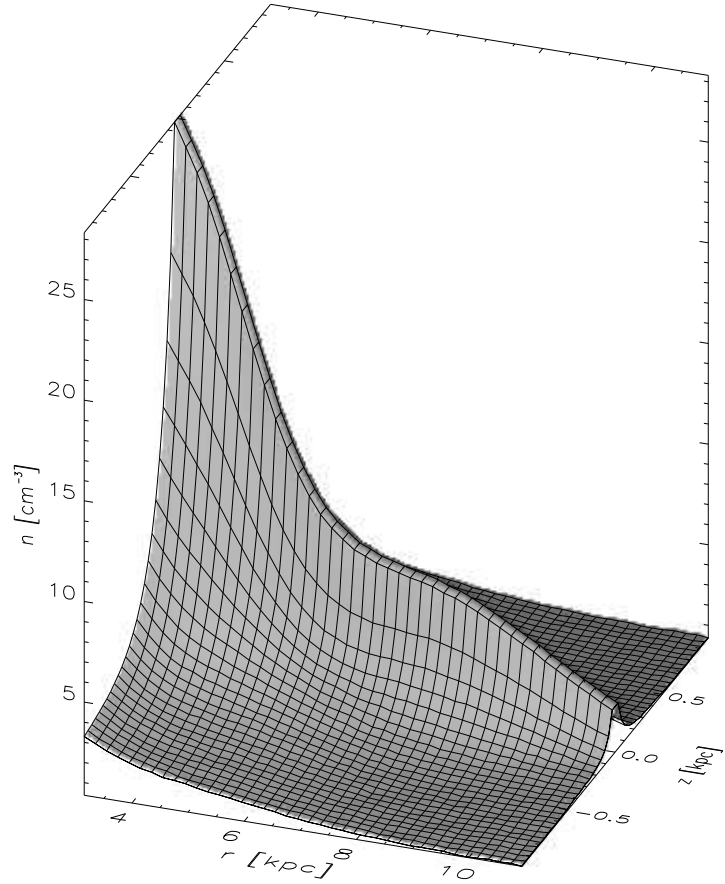}
\epsscale{1.00}
\caption{Full Model Density.  Figure shows the full density represented by Model 2 of \citet{D&B} plus a sinusoidal arm perturbation with our standard parameters (see Section \ref{section:Exam}) with an amplitude at 8 kpc of 57\% of the thin disk density.  This is the two-armed spiral model used by \citet{G&C} to explore the gaseous response in MHD.  Note that the arm/interarm contrast is small even at this amplitude.}
\label{fig16}
\end{figure}

The exact density distribution reflected in the $n$th term of the potential of Equation~\ref{equation:Eight} is given by:
\begin{eqnarray*}
\rho_{\rm n}(r, \phi, z) &=& {\rm C_n} \rho_{\rm o} \left( \frac {H}{D_{\rm n} r} \right)  \exp \left( -\frac {r-r_{\rm o}}{R_{\rm s}} \right) \left[{\rm sech}\left( \frac {K_{\rm n} z}{\beta_{\rm n}} \right) \right] ^{\beta_{\rm n}} \\
     & & \bigg\{ \left[ (K_{\rm n} r) \left( \frac {\beta_{\rm n} +1}{\beta_{\rm n}} \right) {\rm sech}^{2}\left( \frac {K_{\rm n} z}{\beta_{\rm n}} \right)  -     \frac {1}{K_{\rm n} r} (E^2 + r E^{\prime}) \right] \cos(n \gamma) - 2 E \cos(\alpha)\sin(n \gamma) \bigg\}
%E^{\prime}) \right] \cos(n \gamma) \\
 %     & & - 2 E \cos(\alpha)\sin(n \gamma) \bigg\}
\label{equation:AOne} 
\end{eqnarray*}
where
\begin{eqnarray*}
E &=& 1 + \frac {K_{\rm n} H}{D_{\rm n}} \left[ 1 - \frac {0.3}{(1+0.3 K_{\rm n}H)^2} \right] -  \frac{r}{ R_{\rm s}}\\
      & &  -( K_{\rm n}H) (1+0.8 K_{\rm n}H) \ln \left[ {\rm sech}\left( \frac {K_{\rm n} z}{\beta_{\rm n}} \right) \right] \\
      & &   - 0.4 ( K_{\rm n}H)^2 \left( \frac {K_{\rm n} z}{\beta_{\rm n}} \right) {\rm tanh}\left( \frac {K_{\rm n} z}{\beta_{\rm n}} \right)
\label{equation:ATwo} 
\end{eqnarray*}
and
\begin{eqnarray*}
r E^{\prime} & = & - \frac {K_{\rm n} H}{D_{\rm n}} \left[ 1 - \frac {0.3(1-0.3 K_{\rm n}H)}{(1+0.3 K_{\rm n}H)^3} \right] \\
     & &  + \left[ \frac {K_{\rm n} H}{D_{\rm n}} \left(1 - \frac {0.3}{(1+0.3 K_{\rm n}H)^2} \right) \right]^2 -  \frac{r}{ R_{\rm s}} \\
      & & +( K_{\rm n}H) (1+1.6 K_{\rm n}H) \ln \left[ {\rm sech}\left( \frac {K_{\rm n} z}{\beta_{\rm n}} \right) \right] \\
       & &  - \left[ 0.4 ( K_{\rm n}H)^2 \left( \frac {K_{\rm n} z}{\beta_{\rm n}} \right) {\rm sech}\left( \frac {K_{\rm n} z}{\beta_{\rm n}} \right) \right]^2 \Big/ \beta_{\rm n} \\
     & & + 1.2 ( K_{\rm n}H)^2 \left( \frac {K_{\rm n} z}{\beta_{\rm n}} \right) {\rm tanh}\left( \frac {K_{\rm n} z}{\beta_{\rm n}} \right).
\label{equation:AThree} 
\end{eqnarray*}
We have stated in the text that, except at very small $r$\ or large $z$, the terms other than that in Equation~\ref{equation:Ten} are very small.  Trying to state this more carefully, we note that for $K_n H \gg 1$, $E$\ and $rE^{\prime}$\ approach $K_n z$\ and $-K_n z$, respectively.  In that limit, the density function approaches
\begin{eqnarray*}
\rho_{\rm n}(r, \phi, z) &\approx& {\rm C_n} \rho_{\rm o}   \exp \left( -\frac {r-r_{\rm o}}{R_{\rm s}} \right) \left[{\rm sech}\left( \frac {K_{\rm n} z}{\beta_{\rm n}} \right) \right] ^{\beta_{\rm n}} \\
     & & \bigg\{ \left[{\rm sech}^{2}\left( \frac {K_{\rm n} z}{\beta_{\rm n}} \right)  -     \left(\frac {z}{ r}\right)^2\right] \cos(n \gamma) - 2 \left(\frac {z}{ r}\right) \cos(\alpha)\sin(n \gamma) \bigg\}.
\label{equation:AFour} 
\end{eqnarray*}
From this it is clear that when
\begin{displaymath}
2 \left(\frac {z}{ r}\right) \cos(\alpha) \ge {\rm sech}^{2}\left( \frac {K_{\rm n} z}{\beta_{\rm n}} \right)
\end{displaymath}
the third term will exceed the first.  This occurs even with our standard parameters; its effect can be clearly seen in Figures~\ref{fig3}, \ref{fig5}, and \ref{fig11}, as a shift in phase of the density perturbation above about 500 pc.  So, it is inaccurate to say that the small terms are everywhere negligible compared to the leading term.  What is true is that the small terms will dominate at high $z$\ and small $r$, but in practical cases of moderate $z/r$\ and $K_n H \not> 1$, the small terms will be small where the density is large, and not very noticeable where it is small.  They drop off slowly with $z$, about like the perturbation potential.

\clearpage
\newpage
%
%\begin{figure}
%\plotone{SpiralFig17R.eps}
%\caption{Logarithm of Midplane Density of Disk Plus Arms for 0.2 Amplitude Concentrated Arm 
%Case.  The projected surface density of disk stars is proportional to this midplane density.  There 
%are ten contours per decade.}
%\label{fig17}
%\end{figure}
%

%

\end{document}